\documentclass{article}
\usepackage{arxiv}
\usepackage{graphicx}
\usepackage{microtype}
\usepackage[utf8]{inputenc}
\usepackage[T1]{fontenc}
\usepackage[english]{babel}
\usepackage{dsfont}
\usepackage{amssymb,bm}
\usepackage{amsfonts}
\usepackage{amsmath,etoolbox}
\usepackage{nicefrac}
\usepackage{pifont}
\usepackage{mathtools}
\usepackage{chemarrow}
\usepackage{graphicx}
\usepackage{subcaption}
\usepackage[section]{placeins}
\usepackage{booktabs}
\usepackage{adjustbox}
\usepackage{multirow}
\usepackage{makecell}
\usepackage{enumitem}
\usepackage{hyperref}
\usepackage{url}
\usepackage{xcolor}

\usepackage[capitalise]{cleveref}
\usepackage{doi}
\usepackage{natbib}

\title{Importance of Aligning Training Strategy with Evaluation for Diffusion Models\\in 3D Multiclass Segmentation}
\date{} 					

\author{
\href{https://orcid.org/0000-0002-1184-7421}{\includegraphics[scale=0.06]{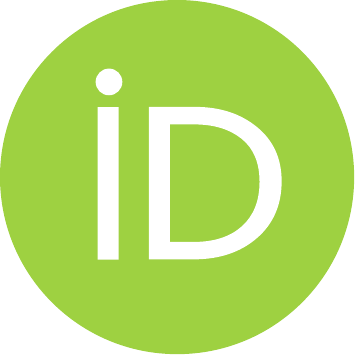}\hspace{1mm}
Yunguan Fu}\\
University College London\\
InstaDeep\\
\And
\href{https://orcid.org/0000-0002-7794-9391}{\includegraphics[scale=0.06]{orcid.pdf}\hspace{1mm}
Yiwen Li} \\
University of Oxford\\
\And
\href{https://orcid.org/0000-0002-5004-0663}{\includegraphics[scale=0.06]{orcid.pdf}\hspace{1mm}
Shaheer U. Saeed} \\
University College London\\
\And
\href{https://orcid.org/0000-0002-5565-1252}{\includegraphics[scale=0.06]{orcid.pdf}\hspace{1mm}
Matthew J. Clarkson
} \\
University College London\\
\And
\href{https://orcid.org/0000-0003-4902-0486}{\includegraphics[scale=0.06]{orcid.pdf}\hspace{1mm}
Yipeng Hu} \\
University College London\\
University of Oxford\\
}

\renewcommand{\shorttitle}{Diffusion Models in 3D Multiclass Segmentation}

\begin{document}
\begin{sloppypar}
\maketitle

\begin{abstract}
Recently, denoising diffusion probabilistic models (DDPM) have been applied to image segmentation by generating segmentation masks conditioned on images, while the applications were mainly limited to 2D networks without exploiting potential benefits from the 3D formulation. 
In this work, we studied the DDPM-based segmentation model for 3D multiclass segmentation on two large multiclass data sets (prostate MR and abdominal CT). We observed that the difference between training and test methods led to inferior performance for existing DDPM methods. To mitigate the inconsistency, we proposed a recycling method which generated corrupted masks based on the model's prediction at a previous time step instead of using ground truth. The proposed method achieved statistically significantly improved performance compared to existing DDPMs, independent of a number of other techniques for reducing train-test discrepancy, including performing mask prediction, using Dice loss, and reducing the number of diffusion time steps during training. The performance of diffusion models was also competitive and visually similar to non-diffusion-based U-net, within the same compute budget.
The JAX-based diffusion framework has been released at \url{https://github.com/mathpluscode/ImgX-DiffSeg}.
\end{abstract}


\section{Introduction}
Multiclass segmentation is one of the most basic tasks in medical imaging, one that arguably benefited the most from deep learning. Although different model architectures~\citep{li2021braingnn,strudel2021segmenter} and training strategies~\citep{fu2019more,li2022prototypical} have been proposed for specific clinical applications, U-net~\citep{ronneberger2015u} trained through supervised training remains the state-of-the-art and an important baseline for many~\citep{ji2022amos}. 
Recently, denoising diffusion probabilistic models (DDPM) have been demonstrated to be effective in a variety of image synthesis tasks~\citep{ho2020denoising}, which can be further guided by a scoring model to generate conditioned images~\citep{dhariwal2021diffusion} or additional inputs~\citep{ho2022classifier}.
These generative modelling results are followed by image segmentation, where the model generates segmentation masks by progressive denoising from random noise. During training, DDPM is provided with an image and a noise-corrupted segmentation mask, generated by a linear interpolation between the ground-truth and a sampled noise. The model is then tasked to predict the sampled noise~\citep{amit2021segdiff,kolbeinsson2022multi,wolleb2022diffusion,wu2022medsegdiff} or the ground-truth mask~\citep{chen2022generalist}.

However, existing applications have been mainly based on 2D networks and, for 3D volumetric medical images, slices are segmented before obtaining the assembled 3D segmentation. Challenges are often encountered for 3D images. 
First, the diffusion model requires image and noise-corrupted masks as input, leading to an increased memory footprint resulting in limited batch size and potentially excessive training time. For instance, the transformer-based architecture becomes infeasible without reducing model size or image, given clinically or academically accessible hardware with limited memory.
Second, most diffusion models assume a denoising process of hundreds of time steps for training and inference, the latter of which in particular leads to prohibitive inference time (e.g., days on TPUs/GPUs).

This work addresses these issues by aligning training with evaluation processes via recycling. As discussed in multiple studies~\citep{chen2022generalist,young2022sud,kolbeinsson2022multi,lai2023denoising}, noise does not necessarily disrupt the shape of ground truth masks and morphological features may be preserved in noise-corrupted masks during training. By training with recycling (\Cref{fig:diffusion}), the prediction from the previous steps is used as input, i.e. rather than the ground truth used in existing methods, for noisy mask sampling. 
This proposed training process emulates the test process since the input is also from the previous predictions at inference time for diffusion models, without access to ground truth.
Furthermore, this work directly predicts ground-truth masks instead of sampled noise~\citep{wu2022medsegdiff}. This facilitates the direct use of Dice loss in addition to cross-entropy during training, as opposed tp $L_2$ loss on noise. Lastly, instead of denoising with at least hundreds of steps as in most existing work, we propose a five-step denoising process for both training and inference, resorting to resampling variance scheduling~\citep{nichol2021improved}.

With extensive experiments in two of the largest public multiclass segmentation applications, prostate MR (589 images) and abdominal CT images (300 images)~\citep{li2022prototypical,ji2022amos}, we demonstrated a statistically significant improvement (between $0.015$ and $0.117$ in Dice score) compared to existing DDPMs. Compared to non-diffusion supervised learning, diffusion models reached a competitive performance (between $0.008$ and $0.015$ in Dice), with the same computational cost. With high transparency and reproducibility, avoiding selective results under different conditions, we conclude that the proposed recycling strategy using mask prediction setting with Dice loss should be the default configuration for 3D segmentation applications with diffusion models.
We release the first unit-tested JAX-based diffusion segmentation framework at \url{https://github.com/mathpluscode/ImgX-DiffSeg}.

\section{Related Work}
The diffusion probabilistic model was first proposed by ~\citet{sohl2015deep} as a generative model for image sampling with a forward noising process. \citet{ho2020denoising} proposed a reverse denoising process that estimates the sampled error, achieving state-of-the-art performance in unconditioned image synthesis at the time.
Different conditioning methods were later proposed to guide the sampling process toward a desired image class or prompt text, using gradients from an external scoring model~\citep{dhariwal2021diffusion,radford2021learning}. Alternatively, \citet{ho2022classifier} showed that guided sampling can be achieved by providing conditions during training.
Diffusion models have been successfully applied in medical imaging applications to synthesise images of different modalities, such as unconditioned lung X-Ray and CT~\citep{ali2023spot}, patient-conditioned brain MR~\citep{pinaya2022brain}, temporal cardiac MR~\citep{kim2022diffusion}, and pathology/sequence-conditioned prostate MR~\citep{saeed2023bi}. The synthesised images have been shown to benefit pre-training self-supervised models~\citep{khader2022medical,saeed2023bi} or support semi-supervised learning~\citep{young2022sud}.

Besides image synthesis, \citet{baranchuk2021label} used pre-trained diffusion models' intermediate feature maps to train pixel classifiers for segmentation, showing these unsupervised models capture semantics that can be extended for image segmentation especially when training data is limited.
Alternatively, \citet{amit2021segdiff} performed progressive denoising from random sampled noise to generate segmentation masks instead of images for microscopic images~\citep{amit2021segdiff}. At each step, the model takes a noise-corrupted mask and an image as input and predicts the sampled noise. Similar approaches have been also applied to thyroid lesion segmentation for ultrasound images~\citep{wu2022medsegdiff} and brain tumour segmentation for MR images with different network architectures~\citep{wolleb2022diffusion,wu2022medsegdiff}.
Empirically, multiple studies~\citep{chen2022generalist,young2022sud,kolbeinsson2022multi} found the noise-corrupted mask generation, via linear interpolation between ground-truth masks and noise, retained morphological features during training, causing potential data leakage. \citet{chen2022generalist} therefore added noises to mask analog bit and tuned its scaling. \citet{young2022sud}, on the other hand, tuned the variance and scaling of added normal noise to reduce information contained in noised masks. Furthermore,~\citet{kolbeinsson2022multi} proposed recursive denoising instead of directly using ground truth for noise-corrupted mask generation.

These works, although using different methods, are all addressing a similar concern: the diffusion model training process is different from its evaluation process, which potentially hinges the efficient learning. Moreover, most published diffusion-model-based segmentation applications have been based on 2D networks. We believe such discrepancy would be more significant when applying 3D networks to volumetric images due to the increased difficulty, resulting in longer training and larger compute cost. In this work, building on these recent developments, we focus on a consistent train-evaluate algorithm for efficient training of diffusion models in 3D medical image segmentation applications. 

\section{Method}
\begin{figure}[!ht]
\begin{center}
    \includegraphics[width=0.9\linewidth]{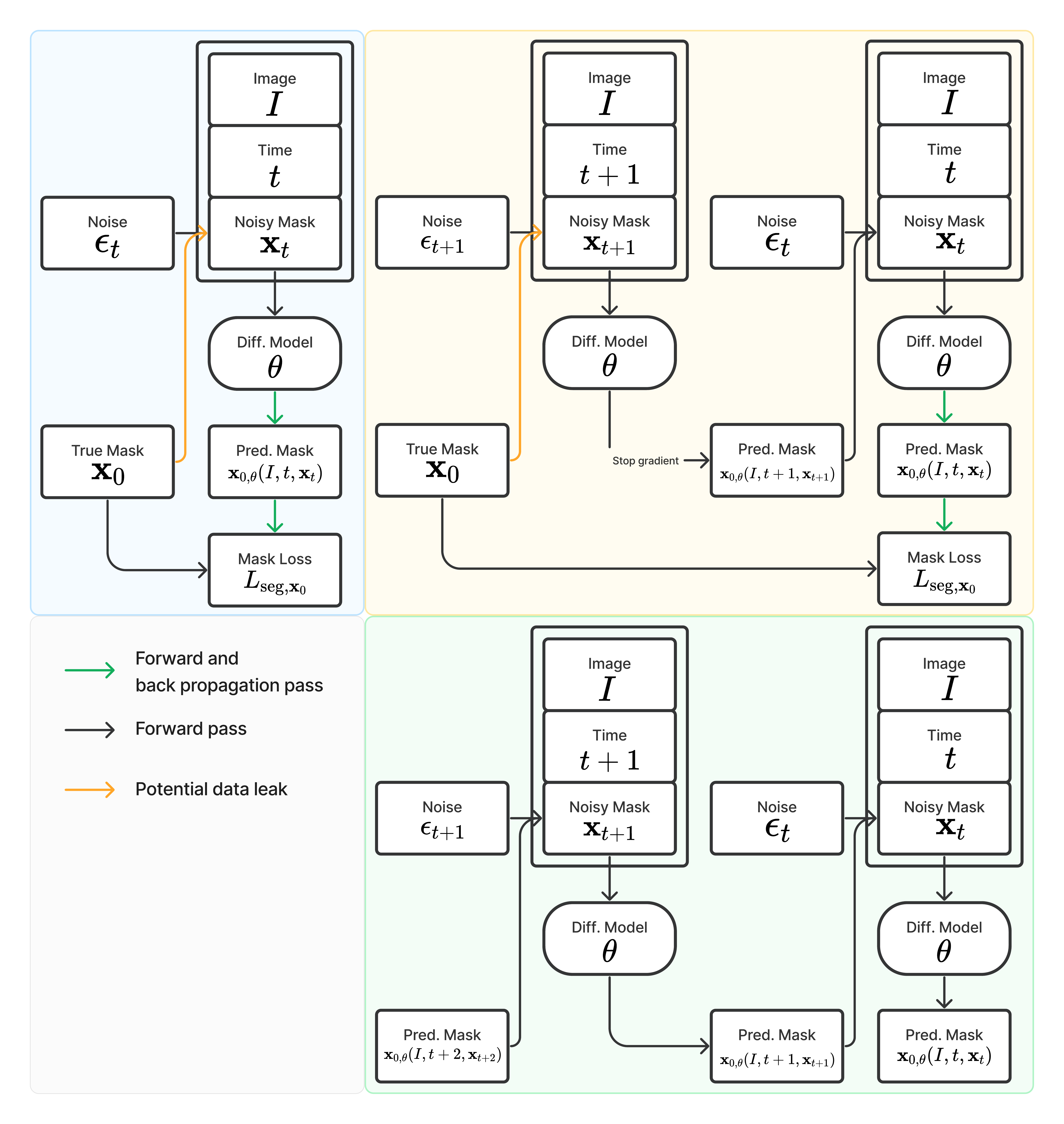}
\caption{Illustration of training with
and without recycling and inference, using mask prediction. For training without recycling (top left), the noisy mask $\mathbf{x}_{t}$ is calculated using the ground truth mask. For training with recycling (top right), $\mathbf{x}_{t}$ is calculated using prediction from the previous step, which is similar to the inference (bottom right).} \label{fig:diffusion}
\end{center}
\end{figure}

\subsection{DDPM for Segmentation}
The denoising diffusion probabilistic models (DDPM)~\citep{sohl2015deep,ho2020denoising,nichol2021improved,kingma2021variational} consider a \textit{forward} process: given sample $\mathbf{x}_0\sim q(\mathbf{x}_0)$, noisy $\mathbf{x}_t$ for $t=1,\cdots,T$ follows a multivariate normal distribution, $q(\mathbf{x}_t\mid\mathbf{x}_{t-1})=\mathcal{N}(\mathbf{x}_t;\sqrt{1-\beta_t}\mathbf{x}_{t-1}, \beta_t\mathbf{I})$,
where $\beta_t\in[0,1]$.
Given a sufficiently large $T$, $\mathbf{x}_T$ approximately follows an isotropic multivariate normal distribution $\mathcal{N}(\mathbf{x}_T;\mathbf{0},\mathbf{I})$. A \textit{reverse} process is then defined to denoise $\mathbf{x}_{t}$ at each step, for $t=T,\cdots,1$, $p_\theta(\mathbf{x}_{t-1}\mid\mathbf{x}_{t})=\mathcal{N}(\mathbf{x}_{t-1};\bm{\mu}_\theta(\mathbf{x}_t,t),\tilde{\beta}_t\mathbf{I})$,
with a predicted mean $\bm{\mu}_\theta(\mathbf{x}_t,t)$ and a variance schedule $\tilde{\beta}_t\mathbf{I}$. $\tilde{\beta}_t=\frac{1 - \bar{\alpha}_{t-1}}{1-\bar{\alpha}_{t}}\beta_t$, where $\alpha_t = 1-\beta_t$, $\bar{\alpha}_t = \prod_{s=0}^t \alpha_s$.
$\bm{\mu}_\theta(\mathbf{x}_{t},t)$ can be modelled in two different ways,
\begin{align}
        \bm{\mu}_\theta(\mathbf{x}_{t},t)&=\frac{\sqrt{\bar{\alpha}_{t-1}}\beta_{t}}{1-\bar{\alpha}_{t}}\mathbf{x}_{0,\theta}(\mathbf{x}_t,t)
    +\frac{1 - \bar{\alpha}_{t-1}}{1-\bar{\alpha}_{t}}\sqrt{\alpha_{t}}\mathbf{x}_{t},~(\text{Predict}~\mathbf{x}_0)\\
    \bm{\mu}_\theta(\mathbf{x}_{t},t)&=\frac{1}{\sqrt{\alpha}_{t}}(\mathbf{x}_{t}
    -\frac{\beta_{t}}{\sqrt{1-\bar{\alpha}_{t}}}\bm{\epsilon}_{t,\theta}(\mathbf{x}_{t},t)),~(\text{Predict noise}~\bm{\epsilon}_t)
\end{align}
where $\mathbf{x}_{0,\theta}(\mathbf{x}_t,t)$ or $\bm{\epsilon}_{t,\theta}(\mathbf{x}_t,t)$ are the learned neural network. 

For segmentation, $\mathbf{x}$ represents a transformed probability with values in $[-1,1]$. Particularly, $\mathbf{x}_0 \in \{1,-1\}$ are binary-valued, transformed from mask. $\mathbf{x}_{0,\theta}$ and $\mathbf{x}_t$ ($t\geq0$) have values in $[-1,1]$. Moreover, the networks $\mathbf{x}_{0,\theta}(I,\mathbf{x}_t,t)$ or $\bm{\epsilon}_{t,\theta}(I,\mathbf{x}_t,t)$ takes one more input $I$, representing the image to segment.

\subsection{Recycling}
During training, existing methods samples $\mathbf{x}_t$ by interpolating noise $\bm{\epsilon}_t$ and ground-truth $\mathbf{x}_0$, which results in a certain level of data leak~\citep{chen2022generalist,young2022sud}. \citet{kolbeinsson2022multi} proposed recursive denoising, which performed $T$ steps on each image progressively to use model's predictions at previous steps. However, this extends the training length $T$ times. Instead, in this work, for each image, the time step $t$ is randomly sampled and the model's prediction $\mathbf{x}_{0,\theta}$ from the previous time step $t+1$ is recycled to replace ground-truth (\cref{fig:diffusion}). 

Similar reuses of the model's predictions have been previously applied in 2D image segmentation~\citep{chen2022generalist}, however $\mathbf{x}_{0,\theta}$ was fed into the network along with $\mathbf{x}_t$ which requires additional memories and still has data leak risks. A further difference to these previous approaches is that, rather than stochastic recycling, usually with a probability of $50\%$, it is always applied throughout the training (which was empirically found to lead to more stable and performant model training). 
Formally, the recycling technique at a sampled step $t$ is as follows,
\begin{align}\label{eq:recycle}
    \mathbf{x}_{t+1}&=\sqrt{\bar{\alpha}_{t+1}}\mathbf{x}_0+\sqrt{1-\bar{\alpha}_{t+1}}\bm{\epsilon}_{t+1},~\text{(Noise mask generation for}~t+1)\\
    \mathbf{x}_{0,\theta}&=\text{StopGradient}(\mathbf{x}_{0,\theta}(I,t+1,\mathbf{x}_{t+1})),~\text{(Mask prediction)}\\
    \mathbf{x}_t&=\sqrt{\bar{\alpha}_t}\mathbf{x}_{0,\theta}+\sqrt{1-\bar{\alpha}_t}\bm{\epsilon}_t,~\text{(Noise mask generation for}~t)
\end{align}
where $\mathbf{x}_{0,\theta}$ is the predicted segmentation mask from $t+1$ using ground-truth, with gradient stopping. $\bm{\epsilon}_t$ and $\bm{\epsilon}_{t+1}$ are two independently sampled noises. Recycling can be applied to models predicting noise (see supplementary materials Section 1 for derivation and illustration).

\subsection{Loss}
Given noised mask $\mathbf{x}_t$, time $t$, and image $I$, the loss can be,
\begin{align}
    L_{\text{seg},\mathbf{x}_0}(\theta) &=\mathbb{E}_{t,\mathbf{x}_0,\bm{\epsilon}_t,I}\mathcal{L}_\text{seg}(
    \mathbf{x}_{0},~\mathbf{x}_{0,\theta}),\\
    L_{\text{seg},\bm{\epsilon}_t}(\theta)
&=\mathbb{E}_{t,\mathbf{x}_0,\bm{\epsilon}_t,I}\|\bm{\epsilon}_t
    -\bm{\epsilon}_{t,\theta}\|_2^2,
\end{align}
where model predict noise $\bm{\epsilon}_{t,\theta}$ or mask $\mathbf{x}_{0,\theta}$ and $\mathcal{L}_\text{seg}$ represents a segmentation-specific loss, such as Dice loss or cross-entropy loss. $t$ is sampled from $1$ to $T$, $\bm{\epsilon}_t\sim\mathcal{N}(\mathbf{0},\mathbf{I})$, and $\mathbf{x}_t(\mathbf{x}_0,\bm{\epsilon}_t)=\sqrt{\bar{\alpha}_t}\mathbf{x}_0+\sqrt{1-\bar{\alpha}_t}\bm{\epsilon}_t$. When model predicts noise, segmentation loss can still be used as the mask can be inferred via interpolation.

\subsection{Variance resampling}
During training or inference, given a variance schedule $\{\beta_t\}_{t=1}^T$ for $T=1000$, a subsequence $\{\beta_k\}_{k=1}^K$ for $K=5$ can be sampled with $\{t_k\}_{k=1}^K$, where $t_K=T$, $t_1=1$,$\beta_k=1-\frac{\bar{\alpha}_{t_k}}{\bar{\alpha}_{t_{k-1}}},
    \tilde{\beta}_{k}=\frac{1 - \bar{\alpha}_{t_k-1}}{1-\bar{\alpha}_{t_k}}\beta_{t_k}$. $\alpha_{k}$ and $\bar{\alpha}_{k}$ are recalculated correspondingly.

\section{Experiment Setting}
\textbf{Prostate MR}~The data set\footnote{\url{https://zenodo.org/record/7013610\#.ZAkaXuzP2rM}}~\citep{li2022prototypical} contains $589$ T2-weighted image-mask pairs for $8$ anatomical structures. The images were randomly split into non-overlapping training, validation, and test sets, with $411$, $14$, $164$ images in each split, respectively. All images were normalised, resampled, and centre-cropped
to an image size of $256\times256\times48$, with a voxel dimension
of $0.75\times0.75\times2.5$ (mm).\\
\textbf{Abdominal CT}~The data set\footnote{\url{https://zenodo.org/record/7155725\#.ZAkbe-zP2rO}}~\citep{ji2022amos} provides $200$ and $100$ CT image-mask pairs for $15$ abdominal organs in training and validation sets. The validation set was randomly split into non-overlapping validation and test sets, with $10$ and $90$ images, respectively. HU values were clipped to $[-991, 362]$ for all images. Images were then normalised, resampled and centre-cropped to an image size of $192\times128\times128$, with a voxel dimension
of $1.5\times1.5\times5.0$ (mm).
\\\textbf{Implementation}~
3D U-nets have four layers with $32$, $64$, $128$, and $256$ channels. For diffusion models, noise-corrupted masks and images were concatenated along feature channels and time was encoded using sinusoidal positional embedding~\citep{rombach2022high}. Random rotation, translation and scaling were adopted for data augmentation during training.
The segmentation-specific loss function is by default the sum of cross-entropy and foreground-only Dice loss. When predicting noise, the $L_2$ loss has a weight of $0.1$~\citep{wu2022medsegdiff}. All models were trained with AdamW for $12500$ steps and a warmup cosine learning rate schedule. Hyper-parameter were configured empirically without extensive tuning.
Binary Dice score and $95\%$ Hausdorff distance in mm (HD), averaged over foreground classes, were reported.
Paired Student's t-tests were performed on Dice score to test statistical significance between model performance with $\alpha=0.01$.
Experiments were performed using bfloat16 mixed precision on TPU v3-8.
The code is available at \url{https://github.com/mathpluscode/ImgX-DiffSeg}.

\section{Results}
\begin{figure}[!ht]
\centering
\includegraphics[width=\linewidth]{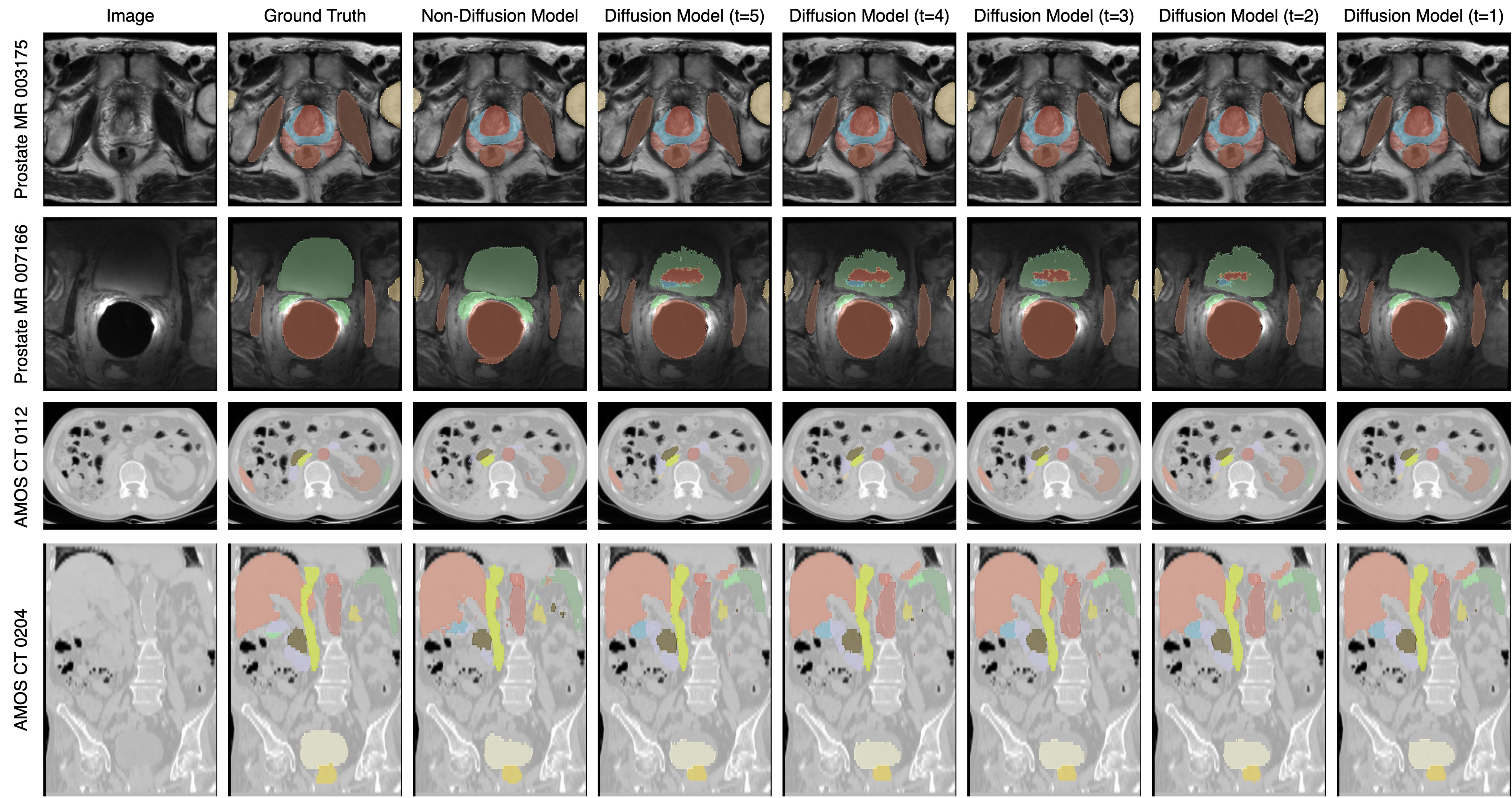}
\caption{Example predictions from diffusion and non-diffusion U-net models.} \label{fig:results}
\end{figure}

\textbf{Recycling}~The performance of the proposed diffusion model is summarised in \cref{tab:recycle}. With recycling, the diffusion-based 3D models reached a Dice score of $0.830$ and $0.801$ which is statistically significantly ($p<0.001$) higher than baseline diffusion with $0.815$ and $0.753$, for prostate MR and abdominal CT respectively. Example predictions were provided in \cref{fig:results}.
\\\textbf{Diffusion vs Non-diffusion}~
The diffusion model is also compared with U-net trained via standard supervised learning in \cref{tab:compare-to-unet}. Within the same computing budget, the diffusion-based 3D model is competitive with its non-diffusion counterpart. The results in \cref{fig:results} are also visually comparable. The difference, however, remains significant with $p<0.001$.
\\\textbf{Ablation studies}~
Comparisons were performed on prostate MR data set for other modifications, including: 
1) predicting mask instead of noise
2) using Dice loss in addition to cross-entropy,
3) using five steps denoising process during training. Improvements in Dice score between $0.09$ and $0.117$ were observed for all modifications (all p-values $<0.001$). The largest improvement was observed when the model predicted segmentation masks instead of noise.
The results were found consistent with the consistency model~\citep{song2023consistency}, which requests diffusion models' predictions of $\mathbf{x}_0$ from different time steps to be similar. Such requirement is implicitly met in our applications as the segmentation loss requires the prediction to be consistent with the ground truth mask given an image. As a result, the predictions from different time steps shall be consistent.
\\\textbf{Limitation}~In general, all methods tend to perform better for large regions of interest (ROI), and there is a significant correlation (Spearman $r>0.8$ and $p<0.01$) between ROI (regions of interest) area and mean Dice score per ROI/class, indicating room of future improvement by addressing small ROIs.

\begin{table}[!ht]
\caption{DDPM with recycling.}
\label{tab:recycle}
\begin{center}
\begin{tabular}{|c|c|c|c|c|}
\hline
\multirow{2}{*}{Recycling} & \multicolumn{2}{c}{Prostate MR} & \multicolumn{2}{|c|}{Abdominal CT} \\ \cline{2-5}
& Dice Score & Hausdorff Dist. & Dice Score & Hausdorff Dist. \\ \hline
N & $0.815\pm0.095$ & $5.485\pm1.069$ & $0.753\pm0.131$ & $9.526\pm2.232$\\ \cline{1-5}
Y &  $0.830\pm0.094$ &$5.424\pm1.176$ & $0.801\pm0.109$ & $9.125\pm2.564$ \\ \hline
\end{tabular}
\end{center}
\end{table}

\begin{table}[!ht]
\caption{Comparison between diffusion models and non-diffusion models.}
\label{tab:compare-to-unet}
\begin{center}
\begin{tabular}{|c|c|c|c|c|}
\hline
\multirow{2}{*}{Diffusion} & \multicolumn{2}{c}{Prostate MR} & \multicolumn{2}{|c|}{Abdominal CT} \\ \cline{2-5}
& Dice score &  Hausdorff Dist. & Dice score &  Hausdorff Dist. \\ \hline
N & $0.838\pm0.088$ & $5.197\pm1.184$ & $0.816\pm0.100$ & $9.091\pm2.475$\\ \cline{1-5}
Y & $0.830\pm0.094$ &$5.424\pm1.176$ & $0.801\pm0.109$ & $9.125\pm2.564$ \\ \hline
\end{tabular}
\end{center}
\end{table}
\begin{table}[!ht]
\caption{Ablation results (prostate MR). HD stands for Hausdorff distance.}
\label{tab:ablation}
\begin{center}
\subfloat[Mask or noise prediction]{
\label{tab:ablation-mask}
\begin{tabular}{|c|c|c|c|}
\hline
Output & Dice Score & HD \\ \hline
Noise & 0.713 & 23.855 \\ \cline{1-3}
Logits & 0.830 & 5.424 \\ \hline
\end{tabular}
}
\subfloat[Dice loss]{
\label{tab:ablation-dice}
\begin{tabular}{|c|c|c|c|}
\hline
Dice Loss & Dice Score & HD \\ \hline
N & 0.812 & 5.463 \\ \cline{1-3}
Y & 0.830 & 5.424 \\ \hline
\end{tabular}
}
\subfloat[\#steps (training)]{
\label{tab:ablation-steps}
\begin{tabular}{|c|c|c|c|}
\hline
Steps & Dice Score & HD \\ \hline
1000 & 0.821 & 5.223 \\ \cline{1-3}
5 & 0.830 & 5.424 \\ \hline
\end{tabular}
}
\end{center}
\end{table}

\section{Discussion}

In this work, we developed a novel denoising diffusion probabilistic model for 3D image multiclass segmentation. By recycling the model's predictions at previous time steps to replace ground truth during training, the method aligns diffusion training and segmentation evaluation, resulting in significant performance improvements compared to existing diffusion methods. Other techniques mitigating training and test inconsistency further improved the diffusion model's performance.
However, the diffusion model did not outperform the non-diffusion-based segmentation models, which have long been well-established.
We believe it is important to report this lack of superior performance in 3D medical image segmentation, especially when experiments are limited to the same compute budget.
Future work could consider other diffusion models such as discrete diffusion and more memory-efficient implementation to enable more complex architectures.
Although the presented experimental results primarily demonstrated methodological development, the fact that these were obtained on two large clinical data sets represents a promising step towards real-world applications. Localising multiple anatomical structures in prostate MR images is key to MR-targeted biopsy, radiotherapy and tissue-preserving focal treatment for patients with urological diseases, while abdominal organ outlines can be directly used in planning gastroenterological procedures and hepatic surgery. 

\section{Acknowledgement}
This work was supported by the EPSRC grant (EP/T029404/1), the Wellcome/EPSRC Centre for Interventional and Surgical Sciences (203145Z/16/Z), the International Alliance for Cancer Early Detection, an alliance between Cancer Research UK (C28070/A30912, C73666/A31378), Canary Center at Stanford University, the University of Cambridge, OHSU Knight Cancer Institute, University College London and the University of Manchester, and Cloud TPUs from Google's TPU Research Cloud (TRC).

\newpage
\bibliographystyle{unsrtnat}
\bibliography{references}

\begin{thebibliography}{28}
\providecommand{\natexlab}[1]{#1}
\providecommand{\url}[1]{\texttt{#1}}
\expandafter\ifx\csname urlstyle\endcsname\relax
  \providecommand{\doi}[1]{doi: #1}\else
  \providecommand{\doi}{doi: \begingroup \urlstyle{rm}\Url}\fi

\bibitem[Li et~al.(2021)Li, Zhou, Dvornek, Zhang, Gao, Zhuang, Scheinost,
  Staib, Ventola, and Duncan]{li2021braingnn}
Xiaoxiao Li, Yuan Zhou, Nicha Dvornek, Muhan Zhang, Siyuan Gao, Juntang Zhuang,
  Dustin Scheinost, Lawrence~H Staib, Pamela Ventola, and James~S Duncan.
\newblock Braingnn: Interpretable brain graph neural network for fmri analysis.
\newblock \emph{Medical Image Analysis}, 74:\penalty0 102233, 2021.

\bibitem[Strudel et~al.(2021)Strudel, Garcia, Laptev, and
  Schmid]{strudel2021segmenter}
Robin Strudel, Ricardo Garcia, Ivan Laptev, and Cordelia Schmid.
\newblock Segmenter: Transformer for semantic segmentation.
\newblock In \emph{Proceedings of the IEEE/CVF international conference on
  computer vision}, pages 7262--7272, 2021.

\bibitem[Fu et~al.(2019)Fu, Robu, Koo, Schneider, van Laarhoven, Stoyanov,
  Davidson, Clarkson, and Hu]{fu2019more}
Yunguan Fu, Maria~R Robu, Bongjin Koo, Crispin Schneider, Stijn van Laarhoven,
  Danail Stoyanov, Brian Davidson, Matthew~J Clarkson, and Yipeng Hu.
\newblock More unlabelled data or label more data? a study on semi-supervised
  laparoscopic image segmentation.
\newblock \emph{arXiv preprint arXiv:1908.08035}, 2019.

\bibitem[Li et~al.(2022)Li, Fu, Gayo, Yang, Min, Saeed, Yan, Wang, Noble,
  Emberton, et~al.]{li2022prototypical}
Yiwen Li, Yunguan Fu, Iani Gayo, Qianye Yang, Zhe Min, Shaheer Saeed, Wen Yan,
  Yipei Wang, J~Alison Noble, Mark Emberton, et~al.
\newblock Prototypical few-shot segmentation for cross-institution male pelvic
  structures with spatial registration.
\newblock \emph{arXiv preprint arXiv:2209.05160}, 2022.

\bibitem[Ronneberger et~al.(2015)Ronneberger, Fischer, and
  Brox]{ronneberger2015u}
Olaf Ronneberger, Philipp Fischer, and Thomas Brox.
\newblock U-net: Convolutional networks for biomedical image segmentation.
\newblock In \emph{Medical Image Computing and Computer-Assisted
  Intervention--MICCAI 2015: 18th International Conference, Munich, Germany,
  October 5-9, 2015, Proceedings, Part III 18}, pages 234--241. Springer, 2015.

\bibitem[Ji et~al.(2022)Ji, Bai, Yang, Ge, Zhu, Zhang, Li, Zhang, Ma, Wan,
  et~al.]{ji2022amos}
Yuanfeng Ji, Haotian Bai, Jie Yang, Chongjian Ge, Ye~Zhu, Ruimao Zhang, Zhen
  Li, Lingyan Zhang, Wanling Ma, Xiang Wan, et~al.
\newblock Amos: A large-scale abdominal multi-organ benchmark for versatile
  medical image segmentation.
\newblock \emph{arXiv preprint arXiv:2206.08023}, 2022.

\bibitem[Ho et~al.(2020)Ho, Jain, and Abbeel]{ho2020denoising}
Jonathan Ho, Ajay Jain, and Pieter Abbeel.
\newblock Denoising diffusion probabilistic models.
\newblock \emph{Advances in Neural Information Processing Systems},
  33:\penalty0 6840--6851, 2020.

\bibitem[Dhariwal and Nichol(2021)]{dhariwal2021diffusion}
Prafulla Dhariwal and Alexander Nichol.
\newblock Diffusion models beat gans on image synthesis.
\newblock \emph{Advances in Neural Information Processing Systems},
  34:\penalty0 8780--8794, 2021.

\bibitem[Ho and Salimans(2022)]{ho2022classifier}
Jonathan Ho and Tim Salimans.
\newblock Classifier-free diffusion guidance.
\newblock \emph{arXiv preprint arXiv:2207.12598}, 2022.

\bibitem[Amit et~al.(2021)Amit, Nachmani, Shaharbany, and
  Wolf]{amit2021segdiff}
Tomer Amit, Eliya Nachmani, Tal Shaharbany, and Lior Wolf.
\newblock Segdiff: Image segmentation with diffusion probabilistic models.
\newblock \emph{arXiv preprint arXiv:2112.00390}, 2021.

\bibitem[Kolbeinsson and Mikolajczyk(2022)]{kolbeinsson2022multi}
Benedikt Kolbeinsson and Krystian Mikolajczyk.
\newblock Multi-class segmentation from aerial views using recursive noise
  diffusion.
\newblock \emph{arXiv preprint arXiv:2212.00787}, 2022.

\bibitem[Wolleb et~al.(2022)Wolleb, Sandk{\"u}hler, Bieder, Valmaggia, and
  Cattin]{wolleb2022diffusion}
Julia Wolleb, Robin Sandk{\"u}hler, Florentin Bieder, Philippe Valmaggia, and
  Philippe~C Cattin.
\newblock Diffusion models for implicit image segmentation ensembles.
\newblock In \emph{International Conference on Medical Imaging with Deep
  Learning}, pages 1336--1348. PMLR, 2022.

\bibitem[Wu et~al.(2022)Wu, Fang, Zhang, Yang, and Xu]{wu2022medsegdiff}
Junde Wu, Huihui Fang, Yu~Zhang, Yehui Yang, and Yanwu Xu.
\newblock Medsegdiff: Medical image segmentation with diffusion probabilistic
  model.
\newblock \emph{arXiv preprint arXiv:2211.00611}, 2022.

\bibitem[Chen et~al.(2022)Chen, Li, Saxena, Hinton, and
  Fleet]{chen2022generalist}
Ting Chen, Lala Li, Saurabh Saxena, Geoffrey Hinton, and David~J Fleet.
\newblock A generalist framework for panoptic segmentation of images and
  videos.
\newblock \emph{arXiv preprint arXiv:2210.06366}, 2022.

\bibitem[Young et~al.(2022)Young, Dalca, Ferrante, Golland, Fischl, and
  Iglesias]{young2022sud}
Sean~I Young, Adrian~V Dalca, Enzo Ferrante, Polina Golland, Bruce Fischl, and
  Juan~Eugenio Iglesias.
\newblock Sud: Supervision by denoising for medical image segmentation.
\newblock \emph{arXiv preprint arXiv:2202.02952}, 2022.

\bibitem[Lai et~al.(2023)Lai, Duan, Dai, Li, Fu, Li, Qiao, and
  Wang]{lai2023denoising}
Zeqiang Lai, Yuchen Duan, Jifeng Dai, Ziheng Li, Ying Fu, Hongsheng Li,
  Yu~Qiao, and Wenhai Wang.
\newblock Denoising diffusion semantic segmentation with mask prior modeling.
\newblock \emph{arXiv preprint arXiv:2306.01721}, 2023.

\bibitem[Nichol and Dhariwal(2021)]{nichol2021improved}
Alexander~Quinn Nichol and Prafulla Dhariwal.
\newblock Improved denoising diffusion probabilistic models.
\newblock In \emph{International Conference on Machine Learning}, pages
  8162--8171. PMLR, 2021.

\bibitem[Sohl-Dickstein et~al.(2015)Sohl-Dickstein, Weiss, Maheswaranathan, and
  Ganguli]{sohl2015deep}
Jascha Sohl-Dickstein, Eric Weiss, Niru Maheswaranathan, and Surya Ganguli.
\newblock Deep unsupervised learning using nonequilibrium thermodynamics.
\newblock In \emph{International Conference on Machine Learning}, pages
  2256--2265. PMLR, 2015.

\bibitem[Radford et~al.(2021)Radford, Kim, Hallacy, Ramesh, Goh, Agarwal,
  Sastry, Askell, Mishkin, Clark, et~al.]{radford2021learning}
Alec Radford, Jong~Wook Kim, Chris Hallacy, Aditya Ramesh, Gabriel Goh,
  Sandhini Agarwal, Girish Sastry, Amanda Askell, Pamela Mishkin, Jack Clark,
  et~al.
\newblock Learning transferable visual models from natural language
  supervision.
\newblock In \emph{International conference on machine learning}, pages
  8748--8763. PMLR, 2021.

\bibitem[Ali et~al.(2023)Ali, Murad, and Shah]{ali2023spot}
Hazrat Ali, Shafaq Murad, and Zubair Shah.
\newblock Spot the fake lungs: Generating synthetic medical images using neural
  diffusion models.
\newblock In \emph{Artificial Intelligence and Cognitive Science: 30th Irish
  Conference, AICS 2022, Munster, Ireland, December 8--9, 2022, Revised
  Selected Papers}, pages 32--39. Springer, 2023.

\bibitem[Pinaya et~al.(2022)Pinaya, Tudosiu, Dafflon, da~Costa, Fernandez,
  Nachev, Ourselin, and Cardoso]{pinaya2022brain}
Walter~HL Pinaya, Petru-Daniel Tudosiu, Jessica Dafflon, Pedro~F da~Costa,
  Virginia Fernandez, Parashkev Nachev, Sebastien Ourselin, and M~Jorge
  Cardoso.
\newblock Brain imaging generation with latent diffusion models.
\newblock \emph{arXiv preprint arXiv:2209.07162}, 2022.

\bibitem[Kim and Ye(2022)]{kim2022diffusion}
Boah Kim and Jong~Chul Ye.
\newblock Diffusion deformable model for 4d temporal medical image generation.
\newblock In \emph{Medical Image Computing and Computer Assisted
  Intervention--MICCAI 2022: 25th International Conference, Singapore,
  September 18--22, 2022, Proceedings, Part I}, pages 539--548. Springer, 2022.

\bibitem[Saeed et~al.(2023)Saeed, Syer, Yan, Yang, Emberton, Punwani, Clarkson,
  Barratt, and Hu]{saeed2023bi}
Shaheer~U. Saeed, Tom Syer, Wen Yan, Qianye Yang, Mark Emberton, Shonit
  Punwani, Matthew~J. Clarkson, Dean~C. Barratt, and Yipeng Hu.
\newblock Bi-parametric prostate mr image synthesis using pathology and
  sequence-conditioned stable diffusion.
\newblock \emph{arXiv preprint arXiv:2303.02094}, 2023.

\bibitem[Khader et~al.(2022)Khader, Mueller-Franzes, Arasteh, Han, Haarburger,
  Schulze-Hagen, Schad, Engelhardt, Baessler, Foersch,
  et~al.]{khader2022medical}
Firas Khader, Gustav Mueller-Franzes, Soroosh~Tayebi Arasteh, Tianyu Han,
  Christoph Haarburger, Maximilian Schulze-Hagen, Philipp Schad, Sandy
  Engelhardt, Bettina Baessler, Sebastian Foersch, et~al.
\newblock Medical diffusion--denoising diffusion probabilistic models for 3d
  medical image generation.
\newblock \emph{arXiv preprint arXiv:2211.03364}, 2022.

\bibitem[Baranchuk et~al.(2021)Baranchuk, Rubachev, Voynov, Khrulkov, and
  Babenko]{baranchuk2021label}
Dmitry Baranchuk, Ivan Rubachev, Andrey Voynov, Valentin Khrulkov, and Artem
  Babenko.
\newblock Label-efficient semantic segmentation with diffusion models.
\newblock \emph{arXiv preprint arXiv:2112.03126}, 2021.

\bibitem[Kingma et~al.(2021)Kingma, Salimans, Poole, and
  Ho]{kingma2021variational}
Diederik Kingma, Tim Salimans, Ben Poole, and Jonathan Ho.
\newblock Variational diffusion models.
\newblock \emph{Advances in neural information processing systems},
  34:\penalty0 21696--21707, 2021.

\bibitem[Rombach et~al.(2022)Rombach, Blattmann, Lorenz, Esser, and
  Ommer]{rombach2022high}
Robin Rombach, Andreas Blattmann, Dominik Lorenz, Patrick Esser, and Bj{\"o}rn
  Ommer.
\newblock High-resolution image synthesis with latent diffusion models.
\newblock In \emph{Proceedings of the IEEE/CVF Conference on Computer Vision
  and Pattern Recognition}, pages 10684--10695, 2022.

\bibitem[Song et~al.(2023)Song, Dhariwal, Chen, and
  Sutskever]{song2023consistency}
Yang Song, Prafulla Dhariwal, Mark Chen, and Ilya Sutskever.
\newblock Consistency models.
\newblock \emph{arXiv preprint arXiv:2303.01469}, 2023.

\end{thebibliography}

\newpage
\section*{Appendix}
\begin{figure}[!ht]
\centering
\includegraphics[width=0.9\linewidth]{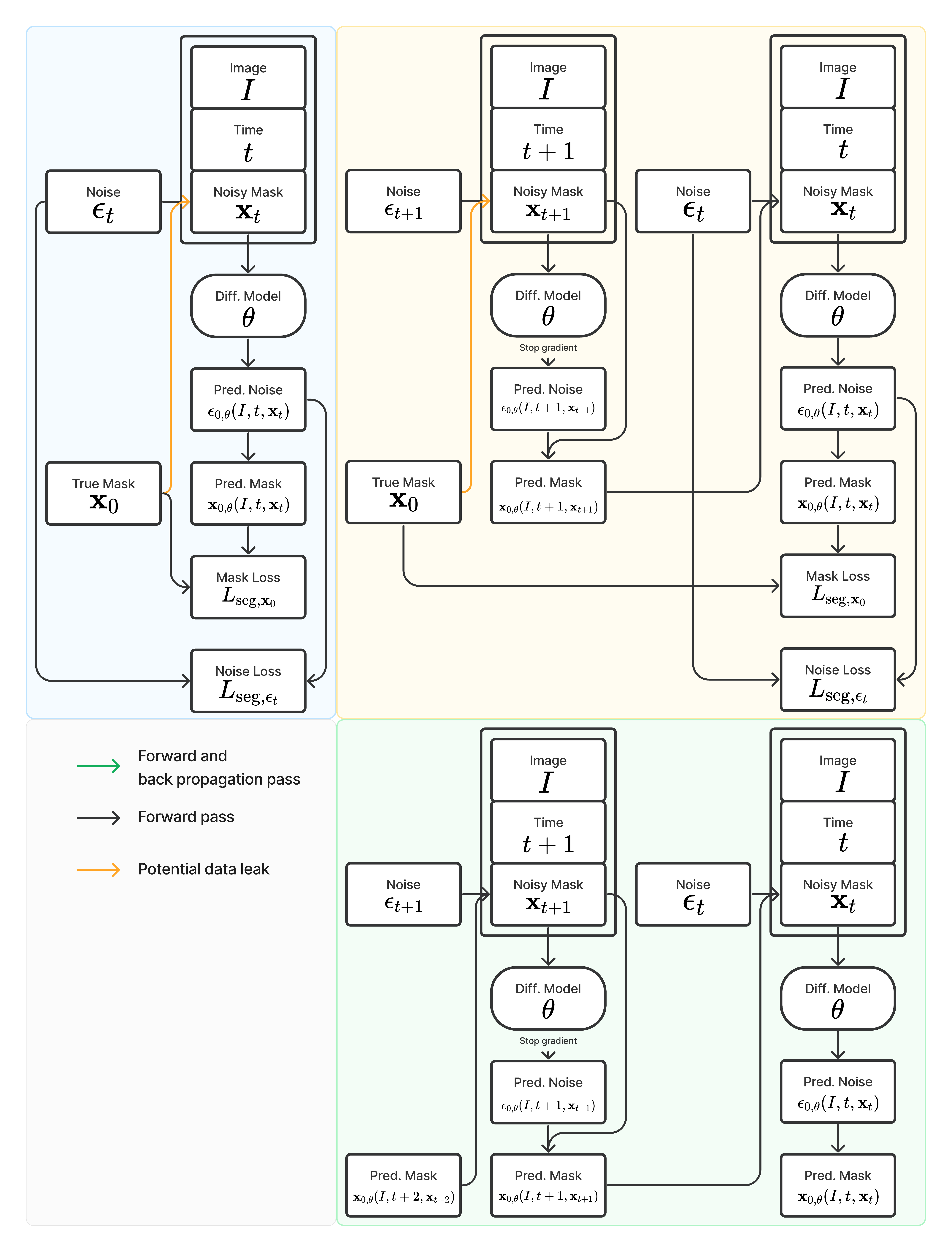}
\caption{Illustration of training with and without recycling, and inference, using noise prediction. For training without recycling (top left), the noisy mask $\mathbf{x}_{t}$ is calculated using the ground truth mask. For training with recycling (top right), $\mathbf{x}_{t}$ is calculated using prediction from the previous step, which is similar to the inference (bottom right).}\label{fig:method_noise}
\end{figure}

Recycling can be applied to models predicting noise as below (\Cref{fig:method_noise}),
\begin{align}\label{eq:recycle-noise}
    \mathbf{x}_{t+1}&=\sqrt{\bar{\alpha}_{t+1}}\mathbf{x}_0+\sqrt{1-\bar{\alpha}_{t+1}}\bm{\epsilon}_{t+1},~\text{(Noise mask generation for}~t+1)\\
    \bm{\epsilon}_{t+1,\theta}&=\text{StopGradient}(\bm{\epsilon}_{t+1,\theta}(I,t+1,\mathbf{x}_{t+1})),~\text{(Noise prediction)}\\
    \mathbf{x}_{0,\theta}&=\frac{1}{\sqrt{\bar{\alpha}_{t+1}}}(\mathbf{x}_{t+1}-\sqrt{1-\bar{\alpha}_{t+1}}\bm{\epsilon}_{t+1,\theta}),~\text{(Mask prediction)}\\
    \mathbf{x}_t&=\sqrt{\bar{\alpha}_t}\mathbf{x}_{0,\theta}+\sqrt{1-\bar{\alpha}_t}\bm{\epsilon}_t.~\text{(Noise mask generation for}~t)
\end{align}
\end{sloppypar}
\end{document}